\documentclass[aps,prl,twocolumn,amsmath,amssymb,superscriptaddress,floatfix,reprint,longbibliography]{revtex4-1}
\usepackage{color}
\usepackage{natbib}
\usepackage[T1]{fontenc}
\usepackage[latin9]{inputenc}
\usepackage{units}
\usepackage{amssymb}
\usepackage{graphicx}
\usepackage{svg}
\usepackage{esint}
\usepackage{bm}
\usepackage{ulem}
\usepackage[colorlinks=true, citecolor=blue]{hyperref}
\setcitestyle{journalcolor= blue}

\begin{document}
\title{Magnetoelectricity induced by rippling of magnetic nanomembranes and wires}

\author{Carmine Ortix}
\affiliation{Dipartimento di Fisica ``E. R. Caianiello", Universit\'{a} di Salerno, IT-84084 Fisciano (SA), Italy}
\email{cortix@unisa.it}
\author{Jeroen van den Brink}
\affiliation{Institute for Theoretical Solid State Physics, IFW Dresden, Helmholtzstr. 20, 01069 Dresden, Germany}

\begin{abstract}
Magnetoelectric crystals have the interesting property that they allow electric fields to induce magnetic polarizations, and vice versa, magnetic fields to generate ferroelectric polarizations. Having such a magnetoelectric coupling usually requires complex types of magnetic textures, e.g., of spiralling type. Here we establish a novel approach to generate a linear magnetoelectric coupling in insulators with a conventional, ferromagnetic ground state. We show that nanoscale curved geometries lead to a reorganization of the magnetic texture that spontaneously breaks inversion symmetry and thereby induces macroscopic magnetoelectric multipoles. Specifically, we prove that structural deformation in the form of controlled ripples activate a magnetoelectric monopole in the recently synthesised two-dimensional magnets. We also demonstrate that in zig-zag shaped ferromagnetic wires in planar architectures, a magnetic toroidal moment triggers a direct linear magnetoelectric coupling.
\end{abstract}
\maketitle

Two-dimensional (2D) atomic crystals are prone to change their physical properties in response to external stimuli, such as strain, and can result in new effective materials in terms of electronic and magnetic properties.  
Electrons in graphene
react to mechanical deformations as if external electromagnetic fields were applied. Strain fields 
realize
effective gauge fields that are opposite in the two graphene valleys. These gauge fields lead to a complete reorganization of the electronic spectrum when they generate a ``pseudo-magnetic" field, {\it i.e.} a magnetic field opposite in the two valleys. The latter leads to the appearance of pseudo-Landau levels~\cite{gui10} which have been directly imaged in graphene nanobubbles~\cite{lev10}, and in flakes supported on nano-pillars~\cite{jia17pseudo}. Strain-induced Landau levels have been also generated in triangular nanoprisms ~\cite{nig19}. A periodic arrangement of pseudomagnetic fields with periods in the tens of nanometer scale and ensuing flat electronic minibands have been instead realized in buckled graphene superlattices~\cite{mao20}. 

A relevant question that arises is whether and how mechanical deformations change the magnetic properties of the recently synthesized atomically thin 2D magnets ~\cite{gon17,hua17,sam17,bur18,gib19}. The main point of this study is to show 
that in analogy with the generation of pseudo-Landau levels in graphene, 
magnetic 2D membranes react to mechanical deformations with 
a peculiar reorganization of the magnetic configuration that leads to the appearance of a specific magnetoelectric multipole: the so-called magnetoelectric monopole~\cite{spa13}. 
Such magnetic state reconfiguration results from the interplay between the local curvature of the 
structure and the magnetic order parameter~\cite{str16,she21,mak22,yan12dw,pac17}
The reorganization of the magnetic state we discuss here applies to
two-dimensional magnets with 
ferromagnetic arrangements and out-of-plane magnetic easy axis, and is of special relevance for ferromagnetic insulators such as CrBr$_3$~\cite{gha18,wan21}, few-layer CrI$_3$~\cite{hua17}, and Cr$_2$Ge$_2$Te$_6$~\cite{gon17}. The presence of a macroscopic magnetoelectric monopole moment then yields a direct linear magnetoelectric coupling. 
We also prove that curvilinear, zig-zag shaped, magnetic wires in planar architectures undergo a geometric-induced reorganization of the magnetic state, which yields
a different magnetoelectric multipole, 
a finite toroidal moment, with a magnetoelectric coupling that is linear as well be it of different symmetry.

We recall that magnetoelectric multipoles are formally defined by considering a 
spin system in a inhomogeneous magnetic field that varies slowly on the scale of the system size~\cite{spa08}. 
The interaction with the magnetic field gradient is then regulated by the tensor ${\mathcal M}_{i j} = \int {\bf r}_i \boldsymbol{\mu}_j({\bf r}) d^3{\bf r}$, with $\boldsymbol{\mu}({\bf r})$ the magnetization density.
This can be decomposed into three irreducible tensors: the pseudoscalar $a={\mathrm tr} {\mathcal M}_{i j} / 3$ defining the magnetoelectric monopole; the toroidal moment~\cite{ede07} dual to the antisymmetric part of the tensor $t_i= \epsilon_{i j k} {\mathcal M}_{j k} / 2$; and the traceless symmetric tensor describing the quadrupole magnetic moment of the system. Being odd under both spatial and time-reversal symmetry, these three irriducible tensors 
directly yield a linear 
coupling  between polarization ${\bf P}$ and magnetization ${\bf M}$. 
In particular, the two linear couplings $A~{\bf P} \cdot {\bf M}$, $A$ being the monopolization $A=a / V$, and ${\bf T} \cdot {\bf P} \times {\bf M}$ with ${\bf T} = {\bf t} / V$ representing the toroidization are entirely symmetry-allowed. 

\begin{figure}[tbp]
    \includegraphics[width=\columnwidth]{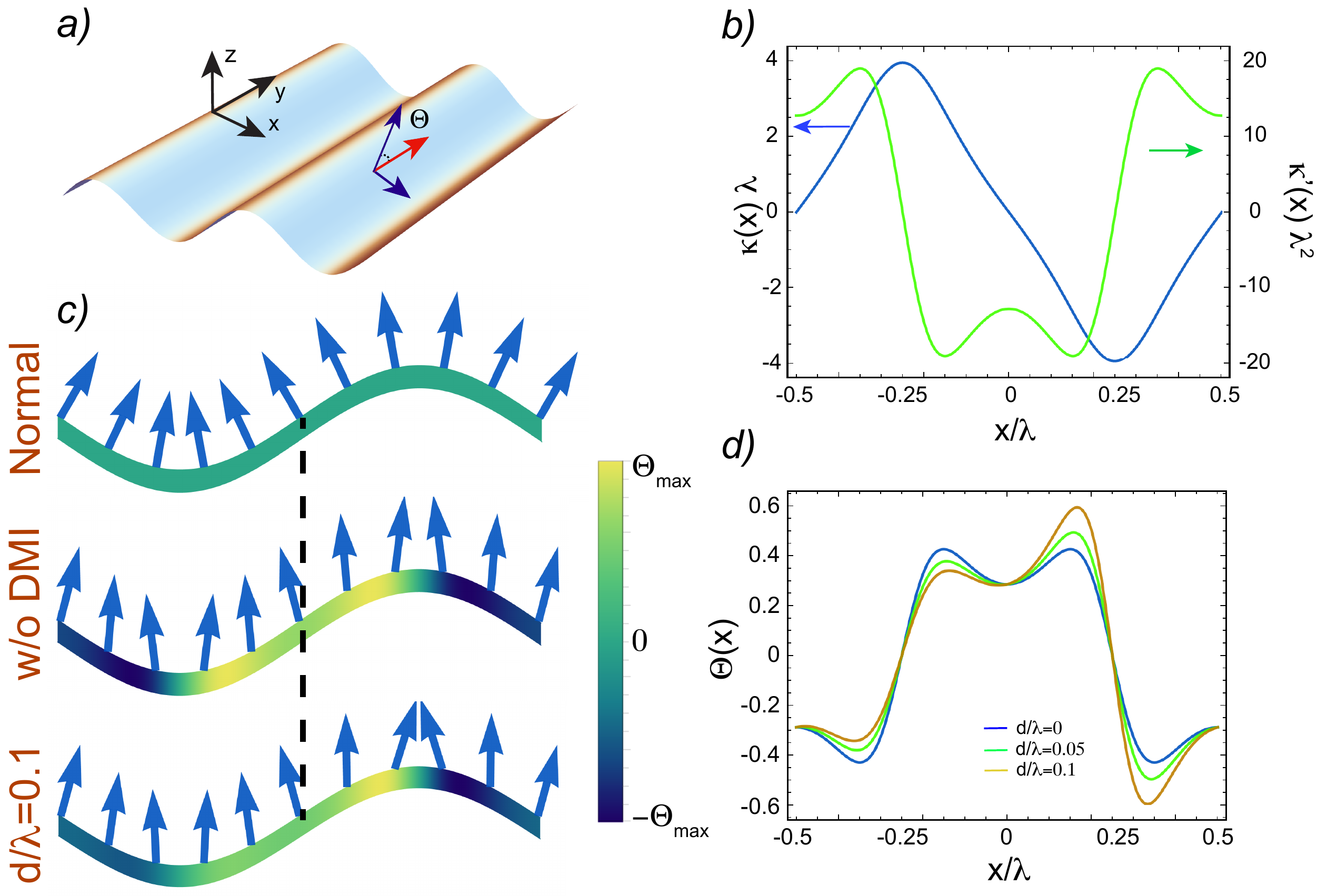}
    \caption{(a) Sketch of a two-dimensional rippled surface with the Euclidean and the local curvilinear coordinates.  $\Theta$ is the canting angle from the normal direction. (b) Behavior  of the local curvature and of its first derivative for a ripple with a sinusoidal shape of height $h=0.1 \lambda$ as a function of the Euclidean coordinate $\hat{x}$. (c) The magnetic ground state excluding curvature effects (top); with curvature effects but in the absence of intrinsic DMI interaction (middle); and in the presence of DMI-induced curvature effects. The density plot shows the local canting angle along the rippled surface. (d) Behavior of the local canting angle of the magnetic texture in the absence ($d/ \lambda = 0$) and presence ($d/\lambda=0.05,0.1$) of intrinsic DMI coupling. The exchange length $l/\lambda=0.15$.}
    \label{fig:fig1}
\end{figure}

To establish how the magnetic ground state of a 2D magnet is affected by geometric deformations, we employ a continuum description that takes into account exchange, magnetocrystalline anisotropy -- this accounts in a local approximation also for magnetostatic interactions --  and an intrinsic DMI coupling. The latter has been shown to be relevant in, for instance,  CrI$_3$~\cite{che18}. 
We consider the deformation to result in one dimensional ripples with a geometry similar to that realized by compressive buckling in graphene on NbSe$_2$~\cite{mao20}. Even if such a buckled layer is locally flat, magneto-mechanical geometric effects still come into play. Specifically, the confinement of the magnetic energy functional to the rippled surface results into an effective curvature-induced DMI coupling~\cite{gai14,kra16} originating from the exchange energy term, and an effective magnetic anisotropy controlled by the local curvature. The latter possesses two contributions of different nature [see Supplemental Material]. First, the intrinsic DMI interaction 
results in an effective anisotropy, which is of the easy-surface type or of the easy-normal (out-of-plane) type depending on the signed curvature [see Supplemental Material]. Second, there is 
an exchange-induced anisotropy that favors an alignment of the magnetic moments along the one-dimensional ripples. 
In magnetic shells with an easy-surface type of magnetocrystalline anistropy, the magnetic ground state is set by this exchange-induced anisotropy [see Supplemental Material]. 
On the contrary, for magnetic layers that in their pristine structure have an out-of-plane ferromagnetic ground state~\cite{gon17,hua17}
the exchange-induced DMI coupling is in full force and leads to inhomogeneous magnetic textures with the magnetic moments lying in the plane perpendicular to the ripples. Even more importantly, the presence of the DMI-induced anisotropy additionally alters the magnetic texture: it is this local change that triggers the appearance of magnetic multipole terms.

To show this, it is convenient to parametrize the direction of the normalized magnetization ${\bf m}={\bf M} / M_s$, with $M_s$ the saturated magnetization, in the locally flat reference frame [see Fig.~\ref{fig:fig1}(a)]. Since, as mentioned above, the magnetic moments lie in the plane perpendicular to the ripple, their direction can be uniquely determined introducing a local angle $\Theta$ that measures the canting of the moments away from the normal direction. Using that the period of the ripple $\lambda$ is much larger than the exchange length $l$ and the DMI length $d$, the canting angle obtained from the minimization of the magnetic energy functional can be expressed [see Supplemental Material] as
$ \Theta \simeq   - \kappa^{\prime}(s)~l^2 / [1 + d~\kappa(s)]$
where $\kappa(s)$ is the local curvature as a function of the arclength $s$ in the corrugated direction, and all lengths have been measured in units of $\lambda$. 
Except for the points of maximum curvature, {\it i.e.} at the crests and valleys of the ripples, the canting angle is generally non-vanishing. The magnetic texture consequently acquires the periodicity of the corrugated structure. The crux of the story is that the presence of the intrinsic DMI coupling results in a crest-valley asymmetry of the magnetic texture with an ensuing magnetically-induced breaking of inversion symmetry. 

\begin{figure}[tbp]
\centering
    \includegraphics[width=\columnwidth]{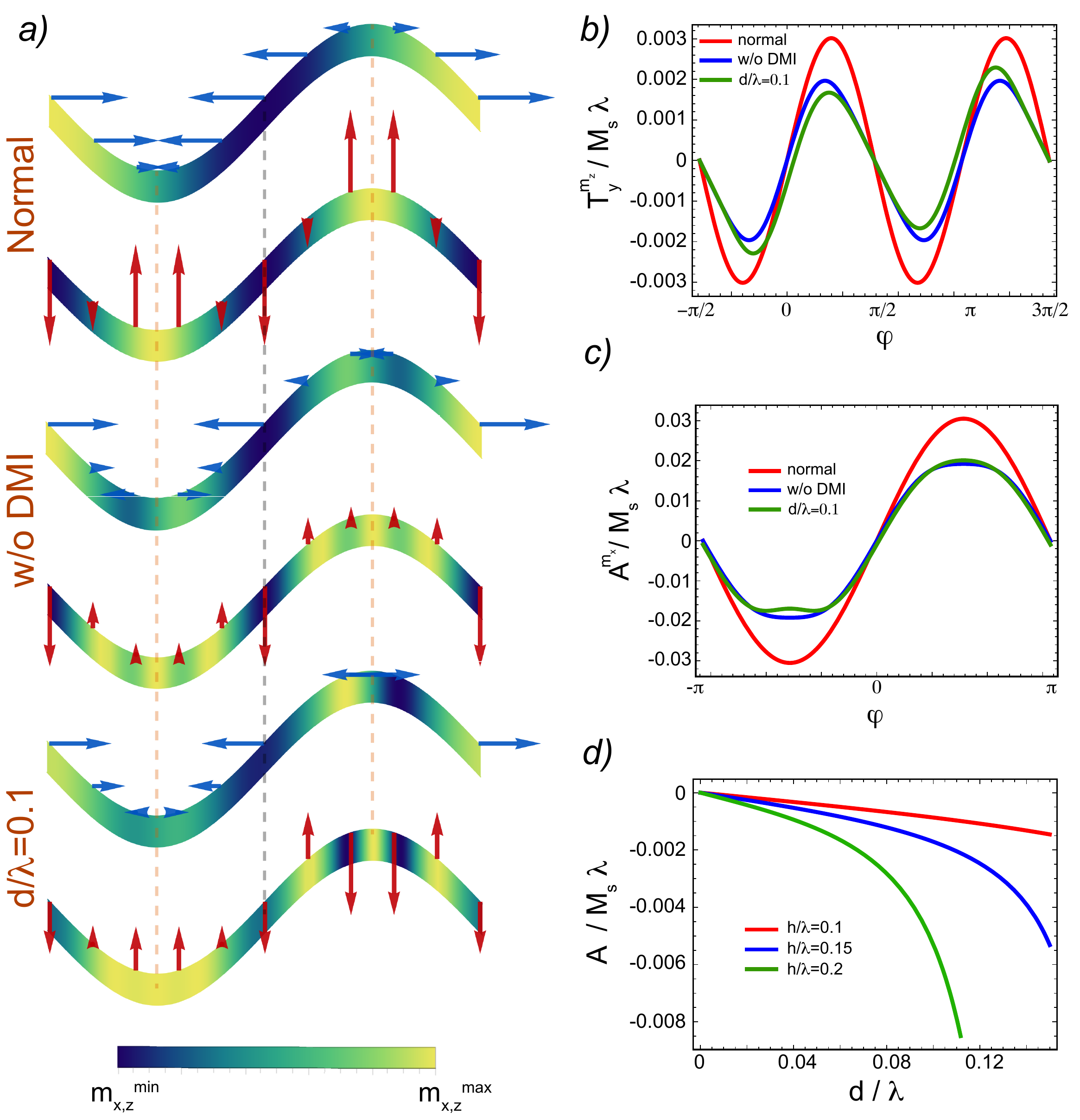}
    \caption{(a) The $\hat{x}$ and ${\hat z}$ components of the compensated magnetic texture  by artificially neglecting curvature effects (top panels), including only exchange-induced curvature effects with $l/\lambda=0.1$ (middle panels), and in presence of the intrinsic DMI coupling with $d/\lambda=l/\lambda$ (bottom panels). The ripple height $h=0.1 \lambda$. The grey dashed line indicates the inversion center of the ripple while the orange lines are the mirror points located at the crests and valleys of the ripple. Same for the phase dependence of the toroidization component $T_y$ (b) and monopolization generated by the $\hat{x}$ component of the magnetization (c). The phase on the sinusoidal ripple profile has been imposed as $z(x)=h \sin{(2 \pi x / \lambda + \varphi)}$. (d) Behavior of the total monopolization as a function of the intrinsic DMI length $d / \lambda$ for ripples of different heights $h$. The exchange length has been set to $l/\lambda=0.1$.}
    \label{fig:fig2}
\end{figure}

Consider for instance a regular periodic wrinkled pattern with a simple sinusoidal shape. In the embedding three-dimensional Euclidean space [c.f. Fig.~\ref{fig:fig1}(a)], this structure is centrosymmetric with the midpoints between the valleys and the crests corresponding to the inversion symmetry centers. Without intrinsic DMI interaction, the canting of the magnetic moments away from the normal direction, and hence the magnetic texture itself, is entirely set by the first derivative of the local curvature, which is even under inversion [c.f. Fig.~\ref{fig:fig1}(b),(c)].
Due to the oddness of the signed curvature, the DMI-induced effective anisotropy introduces a valley-crest asymmetry in the canting angle [c.f. Fig.~\ref{fig:fig1}(c),(d)], and inevitably breaks the inversion symmetry of the magnetic texture. This can be more easily shown by separating the compensated, {\it i.e.} with zero net magnetic moment, parts of the magnetic texture from the uncompensated ferromagnetic one that  preserves the centrosymmetry of the corrugated layer. 
Fig.~\ref{fig:fig2}(a) 
displays the zero-averaged magnetic texture decomposed in its two Euclidean 
components. 
The magnetic texture is 
inversion symmetric both when completely neglecting curvature effects, {\it i.e} for $\Theta \equiv 0$, and when considering only exchanged-induced terms. In addition, it preserves the combined symmetry ${\mathcal M}_x \mathcal{T}$ where $\mathcal{T}$ indicates  time-reversal symmetry while ${\mathcal M}_x$ is the vertical mirror operation with respect to the two mirror planes located at the crests and valleys of the wrinkle. 
A finite DMI breaks the centrosymmetry while still conserving ${\mathcal M}_x \mathcal{T}$. 
Note that ${\mathcal M}_x \mathcal{T}$ constraints the $\hat{z}$ and $\hat{x}$ components of the normalized magnetization to be respectively even and odd with respect to the mirrors located at $x= \pm \lambda / 4$.

Having established that the intrinsic DMI coupling triggers a magnetic inversion symmetry breaking via its curvature-induced effective anisotropy, we next show that the end-product of this phenomenon is the appearance of a non-vanishing magnetoelectric monopole moment. 
We first recall that one of the main complications associated with the definition of higher-order magnetic moments in periodic structures is the fact that, beside the origin dependence characteristic of uncompensated magnetic textures~\cite{ede07,spa13}, they assume arbitrary values depending on the ``unit cell" choice. This is analogous to the situation encountered for the electric polarization in periodic crystals~\cite{kin93,res94}, which, according to the modern theory of polarization, can be only defined modulo a polarization quantum~\cite{vdb93}. In atomic lattices, it has been suggested the existence of toroidization~\cite{ede07} and monopolization~\cite{spa13} quanta, with branch-independent changes in toroidization and monopolization that then acquire physical meaning. 
The geometric superstructures of the present study are assumed to have a period that is order of magnitudes larger than the lattice constant. This then makes the monopolization and toroidization lattices infinitely dense. Nevertheless, we can meaningfully define the magnetic moments in a continuum description using symmetry constraints.
First, and as mentioned above, the uncompensated ferromagnetic $\hat{z}$ component $\overline{m}=\int \sqrt{g}~m_z(x) d x / \int \sqrt{g}~d x$, with $\sqrt{g}$ the line element of the ripple, preserves inversion symmetry. The related magnetoelectric multipoles do not have physical significance. Therefore, we will have to consider only the contributions generated by the compensated magnetic texture. 
These remaining contributions, and this is key, are also subject to constraints imposed by (anti)unitary symmetries. 
Let us consider the $\hat{y}$ component of the toroidization measured in units of $M_s \lambda$ and defined by $T_{y}= \int \sqrt{g} \left[ z(x) m_x(x) - x \left(m_z(x) - \overline{m} \right) \right] d x / \int 2 \sqrt{g} d x$ where $z(x)$ indicates 
the local height of the ripple in the  so-called Monge gauge.
To account for different choices of the supercell, we continuously sweep a phase $\varphi$ on the sinusoidal ripple profile. Fig.~\ref{fig:fig2}(b) displays the corresponding behavior of the toroidization. In the absence of intrinsic DMI coupling ($d=0$) the torodization has odd parity both around $\varphi=0,\pi$, {\it i.e.} for unit cells centered at the inversion centers of the ripples, as well as around $\varphi=\pi/2,3 \pi / 2$, in which case the unit cells are centered at the crests and valleys of the wrinkles. The presence of a finite intrinsic DMI coupling removes the 
parity symmetry around $\varphi=0,\pi$ but keeps the oddness around
$\varphi=\pi/2, 3 \pi/2$. 
This is due to the fact that 
the phase dependence of the toroidization displays the same symmetries of the magnetic system, in strict analogy with the electric polarization lattice~\cite{spa13}. In particular, the odd parity around $\varphi=\pi/2 , 3 \pi/2$ results from the antiunitary ${\mathcal M}_x \mathcal{T}$ symmetry, which is conserved independent of the presence of the intrinsic DMI coupling. Since $T_y \rightarrow -T_y$ under ${\mathcal M}_x \mathcal{T}$, we conclude that the macroscopic toroidization component $T_y$ is forced to vanish. 
In an analogous manner, also the $\hat{x}$ and $\hat{z}$ component of the macroscopic toroidization are vanishing due to the presence of the antiunitary $\mathcal{M}_y \mathcal{T}$ symmetry, which 
is preserved in the system at hand since the magnetic moments are always orthogonal to the translationally invariant $\hat{y}$ direction.

With the macroscopic toroidization that is symmetry forbidden by the ${\mathcal M}_{x , y} {\mathcal T}$ symmetries, we next consider the magnetoelectric monopolization. In Fig.~\ref{fig:fig2}(c) we show the phase dependence of the monopolization associated with the  $\hat{x}$ component of the compensated magnetic texture. 
It has odd-parity around $\varphi=0,\pi$ both when neglecting curvature effects all together and when accounting for exchange-induced terms. This signals the presence of inversion symmetry that is preserved as long as the intrinsic DMI coupling is neglected. With magnetic inversion symmetry breaking ($d \neq 0$) 
the phase dependence does not display any parity, thus implying a finite macroscopic monopolization. Its absolute value is uniquely determined by the fact that its zeroness with inversion symmmetry fixes the ``gauge" $\varphi \equiv 0,\pi$. 
Importantly, also the $\hat{z}$ component of the magnetic texture generates a finite monopolization. This, however, is phase independent for the simple reason that the monopolization local density $\sqrt{g} z(x) m_z(x)  / \int 3 \sqrt{g} d x $ is a periodic function. Fig.~\ref{fig:fig2}(d) shows the ensuing total monopolization as a function of the intrinsic DMI length for ripples of different height $h$. Increasing curvature boosts the monopolization of a rippled two-dimensional ferromagnet. 

\begin{figure}[tbp]
    \includegraphics[width=\columnwidth]{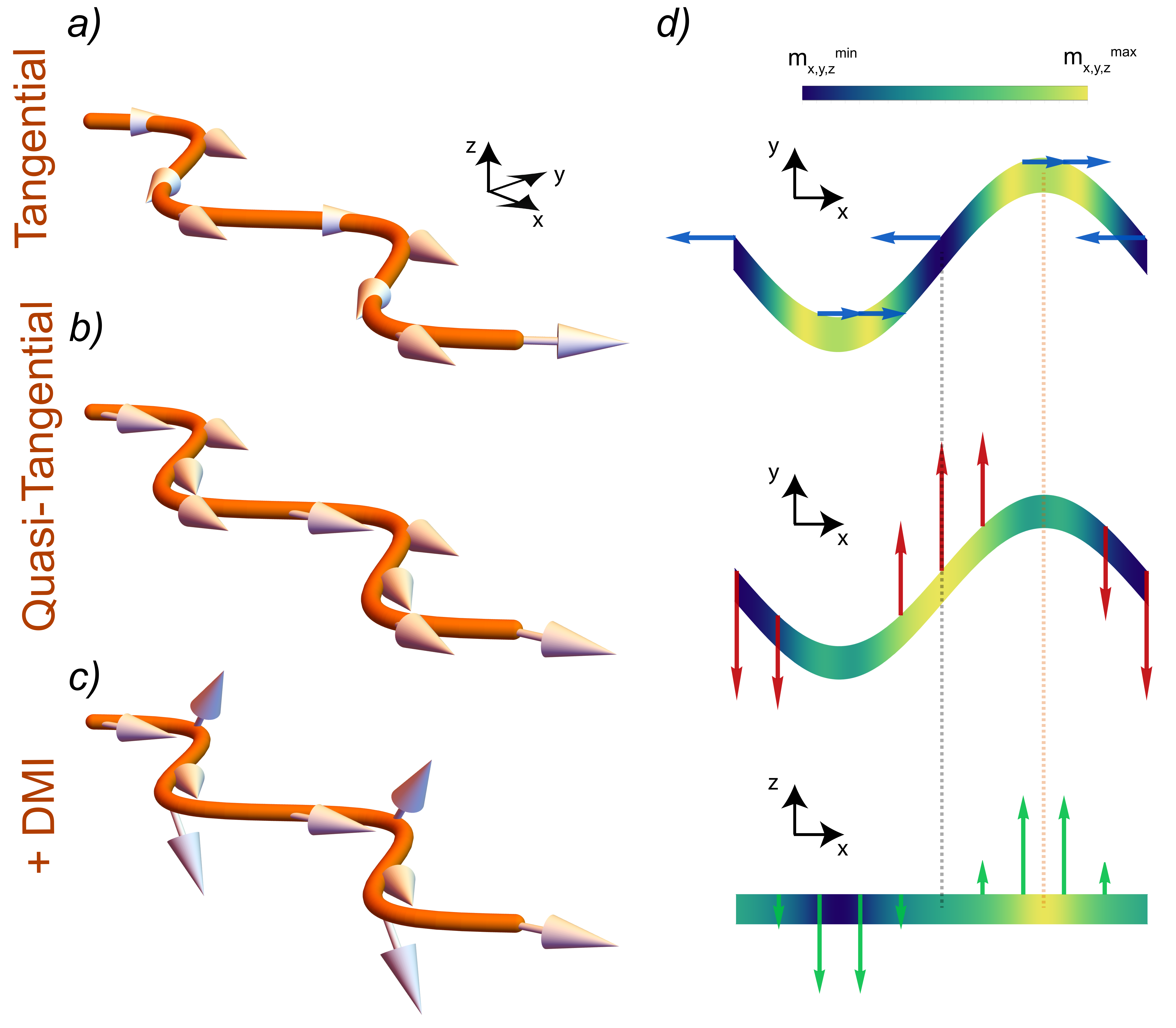}
    \caption{(a),(b),(c) Tangential, quasi-tangential, {\it i.e.} including only exchange-induced curvature effects, and DMI-induced out-of-plane canted magnetic ground state for a magnetic nanowire with zig-zag shape. Panel (d) shows the local behavior of the Euclidean components of the compensated normalized magnetic texture. We also show the inversion and rotation axis centers discussed in the main text.}
    \label{fig:fig3}
\end{figure}

Magnetoelectric multipoles stabilized by the concomitant presence of geometric curvature and intrinsic DMI coupling are not specific of curved two-dimensional membranes. As we now show, zig-zag shaped magnetic wires in planar structures 
can 
also display finite magnetoelectric moments. This is of immediate relevance, for instance, to yttrium iron garnet structures. In thin films, sizable interfacial DMI interaction has been recently reported~\cite{wan20}. 
In analogy with Refs.~\cite{vol18,she22}, we consider in the remainder magnetic wires with a tangential anisotropy and a DMI vector parallel to the tangential direction so that a straight wire has a magnetization parallel to it [see Fig.~\ref{fig:fig3}(a)]. 
In the absence of intrinsic DMI the magnetic state has a characteristic 
quasitangential distribution~\cite{vol18} [see Fig.~\ref{fig:fig3}(b) and the Supplemental Material], 
uniquely identified by the azimuthal angle $\Phi \simeq \kappa^{\prime}(s)~l^2$. A finite intrinsic DMI coupling yields a non-zero
polar angle $\Theta \simeq \kappa(s)~d / 2 $. Due to the intrinsic DMI-induced anisotropy, the magnetic textures then acquire a finite out-of-plane component [see Fig.~\ref{fig:fig3}(c)] whose period is set by the geometric curvature. This out-of-plane component  yields a magnetically-induced inversion symmmetry breaking. To prove this, we assume the zig-zag wire to have a sinusoidal shape. Fig.~\ref{fig:fig3}(d) show the corresponding compensated part of the magnetic texture decomposed into its three Euclidean components. Both the $\hat{x}$ and $\hat{y}$ components are even around the inversion centers $x=0, \lambda$ as a consequence of the parity of the curvature derivative. Being related to the local curvature, the out-of-plane component has opposite parity and thus breaks inversion symmetry. 

\begin{figure}[tbp]
    \includegraphics[width=\columnwidth]{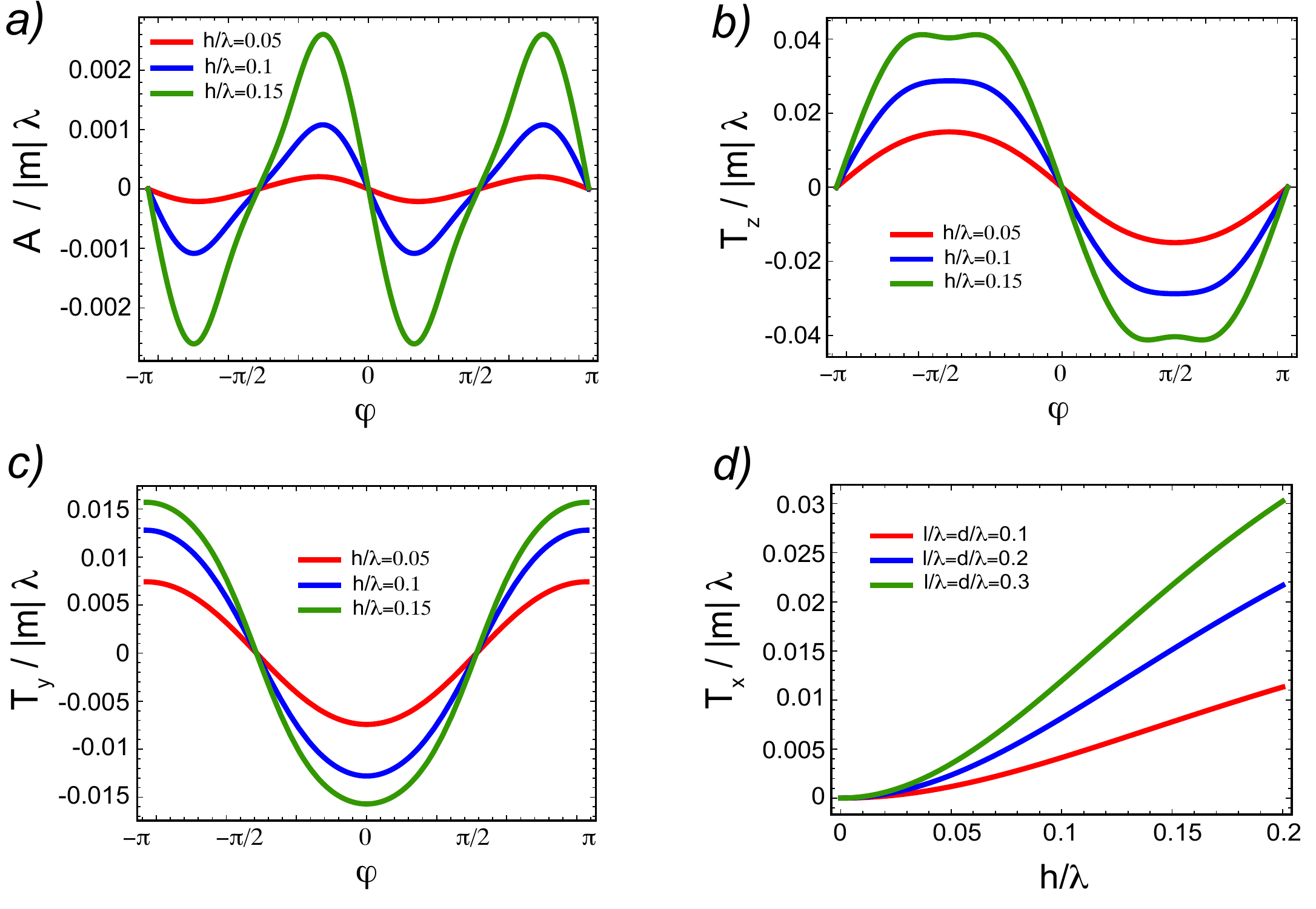}
    \caption{(a),(b),(c) Phase dependence of the monopolization and toroidization components $T_{z,y}$ measured in units of $M_s \lambda$ with $M_s$ the saturated magnetization. We have chosen $l/\lambda = d/ \lambda =0.1$. (d) Behavior of the toroidization as a function of the sinusoidal height $h / \lambda$.}
    \label{fig:fig4}
\end{figure}

However, the complete magnetic ground state preserves two antiunitary symmetries: {\it i)} the combined ${\mathcal C}_{2 y} {\mathcal T}$ where ${\mathcal C}_{2 y}$ is a twofold rotation with a $\hat{y}$-directed rotation axis intersecting the zigzag at its corners. This symmetry regulates the parity of the magnetic texture components (even for $m_{x,z}$ and odd for $m_{y}$)  around the zigzag corners $x=\pm \lambda / 4$; {\it ii)} the combined ${\mathcal C}_{2 z} {\mathcal T}$ where ${\mathcal C}_{2 z}$ is the twofold rotation with an out-of-plane rotation axis intersecting the zigzag at the midpoints between the corners. This symmetry instead regulates the parity of the magnetic textures around $x=0,\lambda/2$. As before, these symmetries pose strong constraints on the presence of the macroscopic monopolization and toroidization. Consider first the monopolization whose phase dependence is shown in Fig.~\ref{fig:fig4}(a). Its odd parity around $\varphi=0,\pi/2$ is consistent with the fact that both ${\mathcal C}_{2 z} {\mathcal T}$ and ${\mathcal C}_{2 y} {\mathcal T}$ send $A \rightarrow - A$. Consequently, the macroscopic monopolization is vanishing. Very similarly, the odd parity of the toroidization component $T_z$ around $\varphi=0$ [see Fig.~\ref{fig:fig4}(b)] and that of $T_y$ around $\varphi=\pi/2$ [see Fig.~\ref{fig:fig4}(c)] are the immediate consequence of the presence of the  ${\mathcal C}_{2 z} {\mathcal T}$ and ${\mathcal C}_{2 y} {\mathcal T}$ symmetry respectively. Since ${\mathcal C}_{2  y,z} {\mathcal T}$ send $T_{y,z} \rightarrow -T_{y,z}$ we conclude that a macroscopic toroidization, if present, has to be directed along the $\hat{x}$ direction. Fig.~\ref{fig:fig4}(d) shows that the macroscopic torodization $T_x$ is finite and increase monotonically both increasing the amplitude of the sinusoidal profile $h$ and the exchange and DMI length. Note that the macroscopic toroidization $T_x$ does not display any phase dependence for the simple reason that its local density $\propto \sqrt{g} y(x) m_z$ is entirely periodic. Therefore, its absolute value can be defined without any ambiguity. 

Curvature effects in two-dimensional ferromagnetic insulators and magnetic wires thus result in the appearance of magnetoelectric multipoles. This phenomenon generally stems from the interplay between the exchange-induced DMI coupling and the intrinsic DMI-induced magnetic anisotropy which lead to a magnetically-induced inversion symmetry breaking. 
Toroidal arrangements with ferrotoroidic domains have been shown to exist in the lithium transition metal phosphatate LiCoPO$_4$~\cite{ake07}. Likewise, LiMnPO$_4$ has been predicted to host a non-vanishing monopolization~\cite{spa13}. Toroidization was also observed in an artificial crystals consisting of planar permalloy nanomagnets~\cite{leu19}. At the nanoscale, magnetoelectric multipoles have been suggested to be present in magnetic insulators with skyrmionic magnetic textures~\cite{bho22}. 
Our study reveals that such multipoles can be designed at the nanoscale starting out even from a ferromagnetic ground state. 
The linear magnetoelectric coupling activated by monopolization and toroidization in these structure represent an important example of geometry-induced magnetic effect at the nanoscale.

\begin{acknowledgements}
We thank D. Makarov, O. Pylypovskyi, K. Yershov and P. Gentile for insightful discussions. C.O. acknowledges support from a VIDI grant (Project 680-47-543) financed by the Netherlands Organization for Scientific Research (NWO). 
\end{acknowledgements}

\end{document}